\documentclass[fleqn,10pt]{wlscirep}
\title{Au-Ag-Cu nano-alloys: tailoring of  permittivity}

\author[1,+]{Yoshikazu Hashimoto}
\author[2,3,+]{Gediminas Seniutinas}
\author[2,3,+]{Armandas Bal\v{c}ytis}
\author[2,3,4]{Saulius Juodkazis}
\author[1,*]{Yoshiaki Nishijima}
\affil[1]{Department of Electrical and Computer Engineering, Graduate School of Engineering, Yokohama National University, 79-5 Tokiwadai, Hodogaya-ku, Yokohama, 240-8501, Japan}
\affil[2]{Centre for Micro-Photonics, Faculty of Engineering and Industrial Sciences, Swinburne
University of Technology, Hawthorn, VIC 3122, Australia}
\affil[3]{Melbourne Centre for Nanofabrication (MCN), Australian National Fabrication Facility, Clayton, VIC 3168, Australia}
\affil[4]{Center for Nanotechnology, King Abdulaziz University, Jeddah 21589, Saudi Arabia}
\affil[*]{nishijima@ynu.ac.jp}

\affil[+]{these authors contributed equally to this work}

\begin{abstract}
Precious metal alloys enables new possibilities to tailor materials for specific optical functions.
Here we present a systematic study of the effects of a nanoscale alloying on the permittivity of Au-Ag-Cu metals at 38 different atomic mixing ratios. The permittivity was  measured and analyzed numerically by applying the Drude model. X-ray diffraction (XRD) revealed  the face centered cubic lattice of the alloys.
Both, optical spectra and XRD results point towards an equivalent composition-dependent electron scattering behavior. Correlation between the fundamental structural parameters of alloys and the resulting optical properties is elucidated. Plasmonic properties of the Au-Ag-Cu alloy nanoparticles were investigated by numerical simulations. Guidelines for designing plasmonic response of nano- structures and their patterns are presented from the material science perspective.
\end{abstract}
\begin{document}

\flushbottom
\maketitle
%
%
\thispagestyle{empty}

\section*{Introduction}

Facile synthesis of novel materials are opening new frontiers in
the key areas of applications, e.g., a recent advances of
perovskite solar cells reaching 20\% efficiency milestone has been
reached just after several years of the basic research\cite{sun,
agiorgousis}. Optical properties of dielectrics, semiconductors
and metals are classified for the tailored absorbance or
refractive index over ultra violet to infrared (UV-IR) spectral
range and are used to design required optical properties of the
layered films\cite{AOM2015}. Recently, it was demonstrated that
amorphous glassy state of mono-atomic metals can be
created\cite{nature}.  This opens new prospectives in alloying and
formation of glass states of metals and tailor their thermal and
electrical conductivity  by control of the structure and atomic
and nanoscale levels.

Currently, alloy nano-materials are being given significant
attention due to emergence of their novel catalytic, energy
storage, and optical functionalities beyond those of pure metals.
It has been demonstrated that solid solutions of Rh and Ag mixed
at the atomic level can store hydrogen much like Pd - a
functionality non-existent in pure forms of Rh and
Ag\cite{kitagawa2}. Another development of great interest is that
Pd-Ru alloy nano-crystals provide for CO oxidization catalysis
superior to that of Rh\cite{kitagawa1}.  Unexpectedly, the Cu-Ag
alloy also catalyzes CO oxidation~\cite{Liu} and also exhibits
anti-bacterial activity\cite{auagalloy}. These new functionalities
are related to the electron binding and charge distribution of the
alloy which alters their chemical behavior, mostly defined by the
electronic properties at the surface~\cite{kitagawa3}.  Also, such
new metal alloys are strongly desired in plasmonic applications
for spectrally tailored optical response. Using an electromagnetic
field enhancement, a more efficient photo-voltaic systems have
been reported~\cite{adhyaksa, sykes, stec}. Highly-doped
semiconductors are utilized as low-loss plasmonic materials, i.e.,
TiN which is complementary metal–oxide–semiconductor
(CMOS)-compatible and can be integrated in plasmonic photonic
circuits~\cite{boltasseva, babicheva, west}. These novel materials
advance the field of functional intermixed metals and
semiconductors beyond the well-established shape memory alloys,
widely used due to their mechanical properties in bio-medical
implant and space applications~\cite{03aem732,Jani}.

Group 11 metals Au, Ag and Cu due to their high conductivity,
chemical stability, thermal conductivity as well as a low thermal
expansion are among the most widely used materials in the field of
electronics. Also, they are prevalent in most plasmonic
applications at the visible light spectrum due to their high
concentration of free electrons and the $d^{10}$ electron
configuration which favors polarizability. Until now, most of the
research on plasmonics has been conducted using pure metals, and
their permittivity has been rigorously measured by means of
ellipsometric studies~\cite{jandc, jandc2,permittivity}; it is
noteworthy, that the permittivity of pure metals is strongly
dependent on the film quality\cite{ACSphotonics}. However,
recently, numerous investigations were launched into plasmonic
applications of dielectrics, semiconductors as well as other kinds
of metals at the electro magnetic (EM) frequencies where damping
losses are significantly lower. As part of this push, the
importance and demand for affordable alloy materials with tunable
properties for plasmonic applications such as surface enhanced
spectroscopy/scattering (SERS) and surface enhanced infrared
absorption (SEIRA) has been growing~\cite{alloyparticle1,
alloyparticle2, alloyparticle3, alloyparticle4, alloyparticle5,
alloyparticle6, ynSEIRA1, ynSEIRA2, ynSERS1, ynSERS2, ynSERS3},
spurred on by prospects in quantitative detection of analytes at
sub-1 ppm levels in the increasingly important fields of
environmental monitoring, indoor air quality control, bio-medical
sensing as well as inspection of chemicals in food, water and
agriculture.

The current lack of knowledge of the complex optical properties of
alloy plasmonic materials calls for direct measurement of relative
permittivities of alloys at different compositions and to compare
these results with analytical predictions obtained using the
effective media theory. In previous work, it was shown how
unreliable an estimate of alloy permittivity based the effective
media formalism could be, even when the permittivity values for
the pure constituent metals are known with high certainty, and how
it is imperative to measure permittivity
experimentally~\cite{ynalloy1, ynalloy2}. Optical, electrical, and
mechanical properties of alloys are strongly dependent on
crystalline structure as experienced by an electron, which in turn
is largely determined by alloying conditions and, therefore, a
wide range of values could be obtained for the same set of
alloying components~\cite{auagalloy1}. As the effective medium
approximations typically neglect possible variations in alloy
structure, systematic investigation over the full range of mixing
ratios (from 0 to 100\% of constituents) is necessary to obtain
meaningful guidelines for predictive design of new plasmonic
materials.

Here, the first study on the experimentally obtained complex
values of permittivity for the $d^{10}$ metals (Au, Ag, and Cu)
alloys, covering the whole intermixing range of possible
stoichiometric variations, deposited by means of thermal
evaporation, is presented. Analysis of the optical properties of
the alloys was carried out within the Drude model framework.
Furthermore, X-ray powder diffraction (XRD) measurements were
carried out in order to reveal crystallinity and intermixing of
the constituent metals, thereby exposing the influence structural
effects have on the plasmonic properties of an alloy.
Finite-difference time-domain (FDTD) numerical simulations, based
on the experimentally defined alloy permittivities, were conducted
for generic nanoparticle-on-a-substrate test structures and
revealed useful guidelines for control over the relaxation time
and spectral position of their localized plasmonic resonances.

\begin{figure}[tb]
\centering\includegraphics[width=12cm]{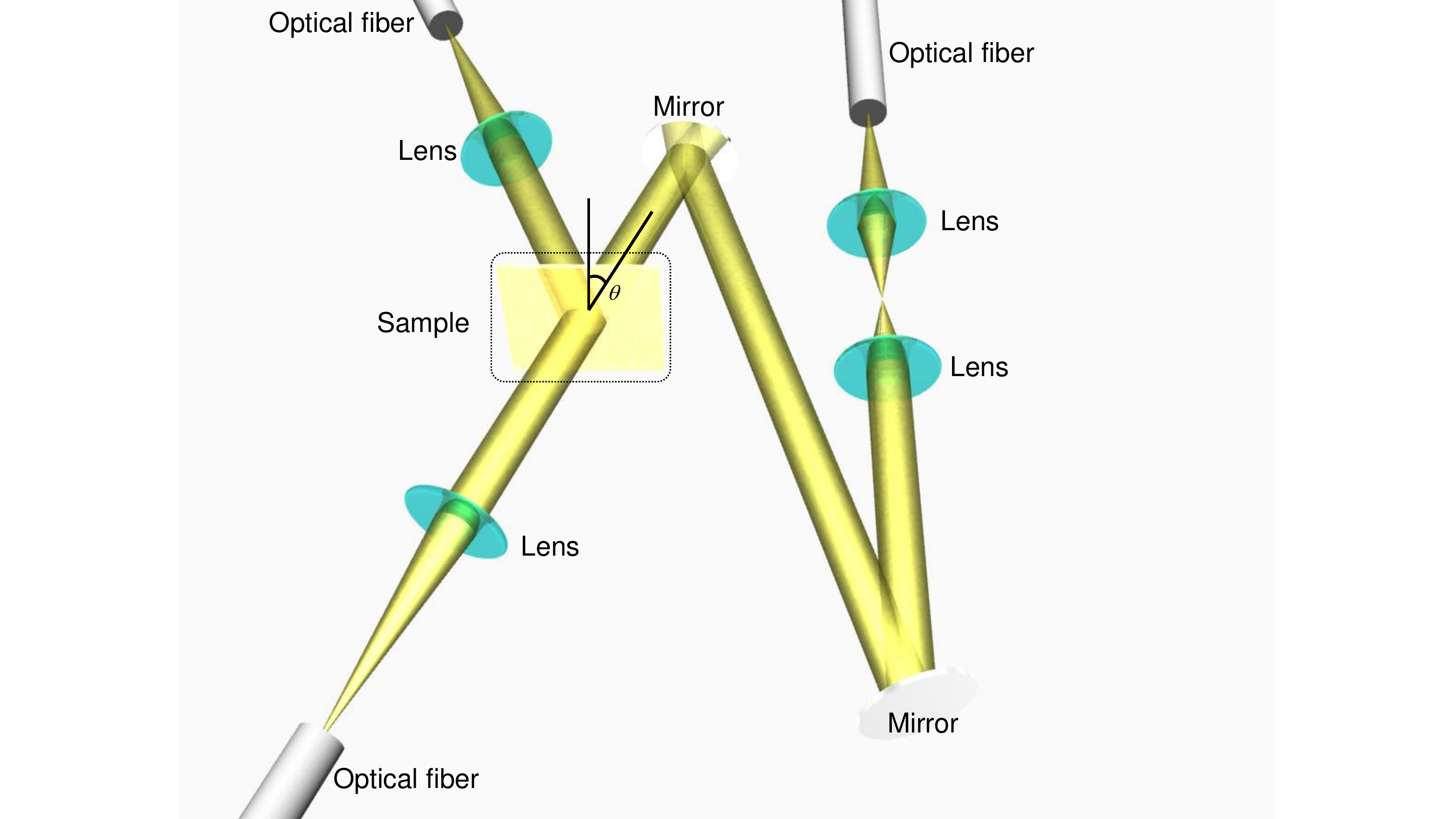} \caption{Layout
of the optical setup used for simultaneous transmittance and
reflectance spectral measurements. Sample was a metal nano-alloy
film on a glass substrate.} \label{f0}
\end{figure}


\begin{figure}[htb]
\centering\includegraphics[width=12cm]{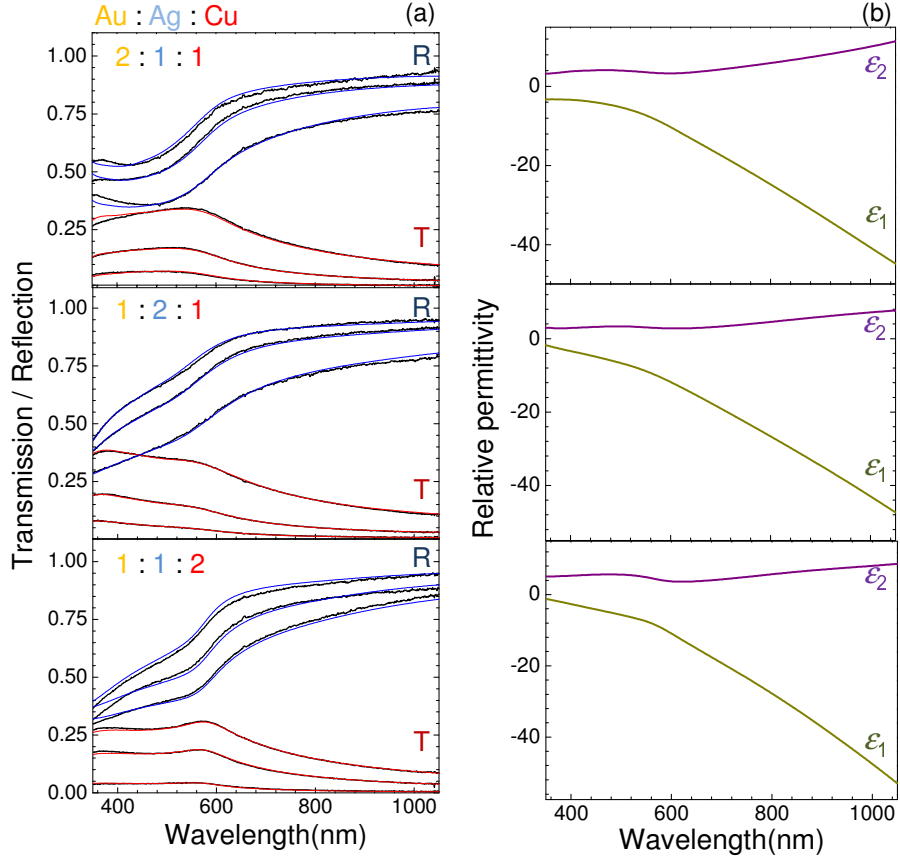} \caption{(a)
Optical transmission, $T$, and reflection, $R$, spectra in the
case of different atomic mixing ratios (Au:Ag:Cu) = (2:1:1),
(1:2:1) and (1:1:2). Three sets of curves each represent layer
thicknesses of 20, 30, and 50~nm, respectively. Red lines denote
the best fit obtained using the Drude-Lorenz model
(Eqn.~\ref{e1}). (b) Spectral dependence of the real,
$\varepsilon_1$, and imaginary, $\varepsilon_2$, parts of the
permittivity of corresponding alloys obtained from the
Drude-Lorenz model analysis. All data analysis was done with $j =
15$ oscillators to achieve high fidelity $F \geq 0.9$ fits. }
\label{f1}
\end{figure}

\section*{Methods}
\subsection*{Samples and optical measurements}

The complex permittivity (the real and imaginary parts) of thin
film alloy layers deposited on glass substrates was determined by
measuring transmission and reflection spectra~\cite{ynalloy2}.
Thin alloy films were prepared using thermal evaporation onto a
glass substrate (Matsunami No.4, 0.35-0.45 mm in thickness).
Chamber pressure was set below 6.0$\times$10$^{-4}$ Pa, evacuated
using a turbo molecular pump. Samples were set onto a planetary
stage and kept at room temperature during deposition with a
sample-to-target distance of 40 cm. The Au, Ag, and Cu targets
were placed onto dedicated tungsten crucibles with a 2 cm
separation. Films with thicknesses of 20, 30, and 50 nm were
deposited at a constant rate of 0.18 nm/min, controlled using a
quartz micro-balance monitor. Alloy layers were formed by
alternating the deposition of high purity Au, Ag and Cu metals
with an alternating step of 1 nm for good intermixing. Schematics
of the optical setup used for measurements of
transmission/reflection spectra are shown in Figure 1.

Fiber coupled halogen deuterium D$_2$ lamp (L10290, Hamamatsu
Photonics Co. Ltd.) was used as the UV-IR light source.
Un-polarized light diffracting from the fiber was collimated into
a 1 cm diameter beam which impinges onto the sample. The
transmitted and reflected light was collected using a
charge-coupled device (CCD) array detector (C10083CA-1050,
Hamamatsu Photonics Co. Ltd.). The experimentally obtained spectra
were analyzed using the FEDataAnalysis permittivity analysis
software (Ohtsuka Electronics Co. Ltd.). The Drude-Lorenz model
was fitted to the experimental data using least-squares method
~\cite{burns}. The free and bound electron derived permittivity is
defined in the following form~\cite{vardeman}:
\begin{equation}\label{e1}
\varepsilon(\omega) = \varepsilon(\infty) -
\frac{\omega_{p}^{2}}{\omega^2 + i\omega\Gamma} + \Sigma_{j}
\frac{A_j \hbar \omega_{0,j}}{(\hbar \omega_{0,j}) - (\hbar
\omega)^2 - i\Gamma_{j} \hbar \omega},
\end{equation}
\noindent where the first two terms represent the Drude free
electron model and the last term is the Lorentz contribution
accounting for the bound electrons participating in interband
transitions. Here $\varepsilon(\infty)$ represents permittivity at
the high frequency limit (infinity), $\omega_{p}$ is the plasma
frequency, $\Gamma$ is the damping constant (= 1/$\tau$), $j$
denotes the index of a Lorentz oscillator (a maximum index of $j =
15$ is assumed for higher accuracy of analysis),  $A_j$,
$\omega_{0,j}$ and $\Gamma_{j}$ are the amplitude, the resonant
frequency and the damping constant of the given oscillator $j$,
respectively,   $\hbar$ is the reduced Plank constant, $i$ is the
imaginary unit, and $\omega$ is the optical cyclic frequency. High
fidelity numerical fits to experimentally measured refractive
index data yield Drude parameters determined with high confidence
bounds, with deviations less than 0.02\% .

\subsection*{Structural X-ray characterization}

The crystallinity and grain size of the resulting alloys were
investigated by means of X-ray powder diffraction (XRD) with
RINT-2500 diffractometer (Rigaku Co.) and applying the 2$\theta$
-- $\theta$ method. The angular resolution and scan rate were set
to 0.02$^{\circ}$ and 0.5$^{\circ}$/min, respectively. All alloy
compositions at different mixing ratios of Au, Ag and Cu, were
shown to be arranged in a face centered cubic (fcc) crystal
lattice. Structural analysis was conducted by focusing on the
diffraction peak from the (111) plane.

\subsection*{FDTD calculations}
\begin{figure}[tb]
\centering\includegraphics[width=12cm]{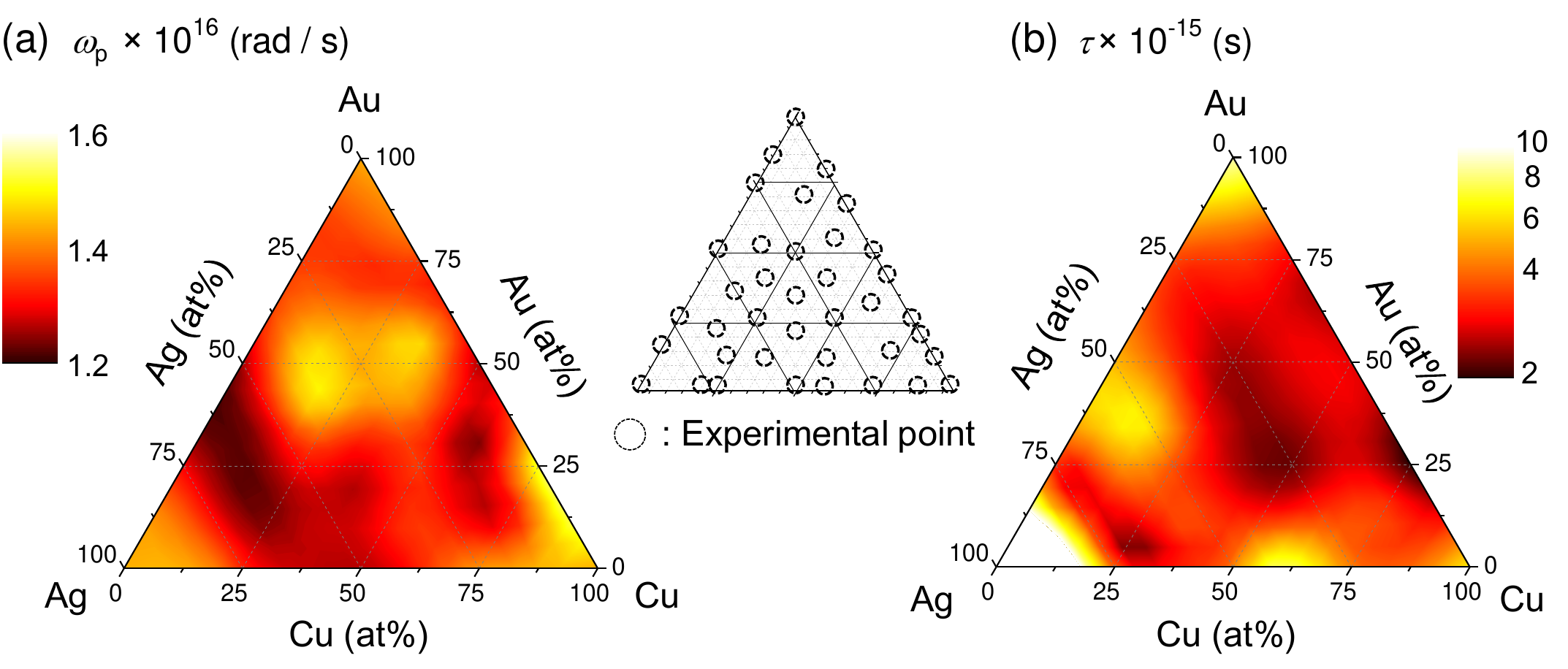} \caption{Interpolated
ternary plots of experimentally obtained Drude parameters (a)
plasma frequency $\omega_{p}$ and (b) relaxation constant $\tau$.
The Au:Ag alloy data were taken from the
reference~\cite{ynalloy1}. Middle inset indicates the atomic
ratios at which experimental measurements were performed.}
\label{f2}
\end{figure}

The finite difference time domain (FDTD) numerical simulations of
typical plasmonic nanostructures were performed using a
commercially available software package FDTD Solutions (Lumerical
Co.) and importing the experimentally determined complex
permittivity values to define the material model. As a test
structure an array of 200-nm-diameter disks of 50 nm thickness on
a SiO$_2$ substrate was modeled. This is a typical plasmonic
structure often fabricated using electron beam lithography (EBL)
and lift-off~\cite{ynSERS1, ynSERS2, ynSERS3, ynrandom1,
ynrandom2}. Periodicity of the disk pattern was set to 500 nm by
using periodic boundary conditions along planes perpendicular to
the substrate plane. The structures were illuminated from the
substrate side using a broadband plane wave source and
frequency-domain field monitors were set at the interface of metal
nanostructures and substrate to obtain electric field distribution
maps for determination of the local field enhancement factors.
Field monitors were set above and below the structures for
transmission and reflection spectral simulations. In order to
obtain the optimal precision, a mesh override region with
$\Delta$x = $\Delta$y = $\Delta$z = 2.5 nm maximum step size was
set in the region surrounding the metallic structure.

\section*{Results and discussion}

Simultaneous measurements of the transmission and reflectance of
thin ``Olympic'' alloy films (Figure 1) were carried out in order
to determine their complex permittivity
$(\varepsilon_1+i\varepsilon_2)$, which in turn defines their
optical properties.

\subsection*{Optical permittivity and Drude parameters}

Optical transmission and reflection spectra were measured for
alloys with 38 different mixing ratios of Au, Ag and Cu, each
deposited as films at thicknesses of 20, 30, and 50 nm.  Examples
of spectra taken at composition ratios of (Au, Ag, Cu) = (2:1:1),
(1:2:1) and (1:1:2) are presented in Figure 2a along with their
best fit to the Drude-Lorenz model using least-squares method (see
the experimental section with more details in the Supplement). The
fitting parameter $F = 1-\sum_{i}^{}{|\Delta Y_{i}|^2}/N_{data}$
was used to check for convergence. Here $N_{data}$ denotes the
experimental data points and $\Delta Y_{i}$ accounts for the
difference between the experimental and simulated values. In this
study, the theoretical model is considered convergent to the
experimental results only when $F\geq 0.90$. Oscillator number of
$j \geq 12$ is required for fitting experimental data for pure Au,
Ag and Cu to be in agreement with permittivity as reported by
Johnson and Christy\cite{jandc}. For alloys it is expected to have
a larger number of possible transitions and up to $j = 15$
oscillators were used to obtain a high fidelity fit. Thereby the
optical permittivity $(\varepsilon_1+i\varepsilon_2)$ in the UV-IR
spectral range has been experimentally determined (Figure 2b) for
different mixing ratios of plasmonic $d^{10}$ family metals with
more details given in the Supplement. High fidelity fits which
were consistently describing reflection and transmission of very
different alloys collaborated validity of the used approach.

\begin{figure}[tb]
\centering\includegraphics[width=12cm]{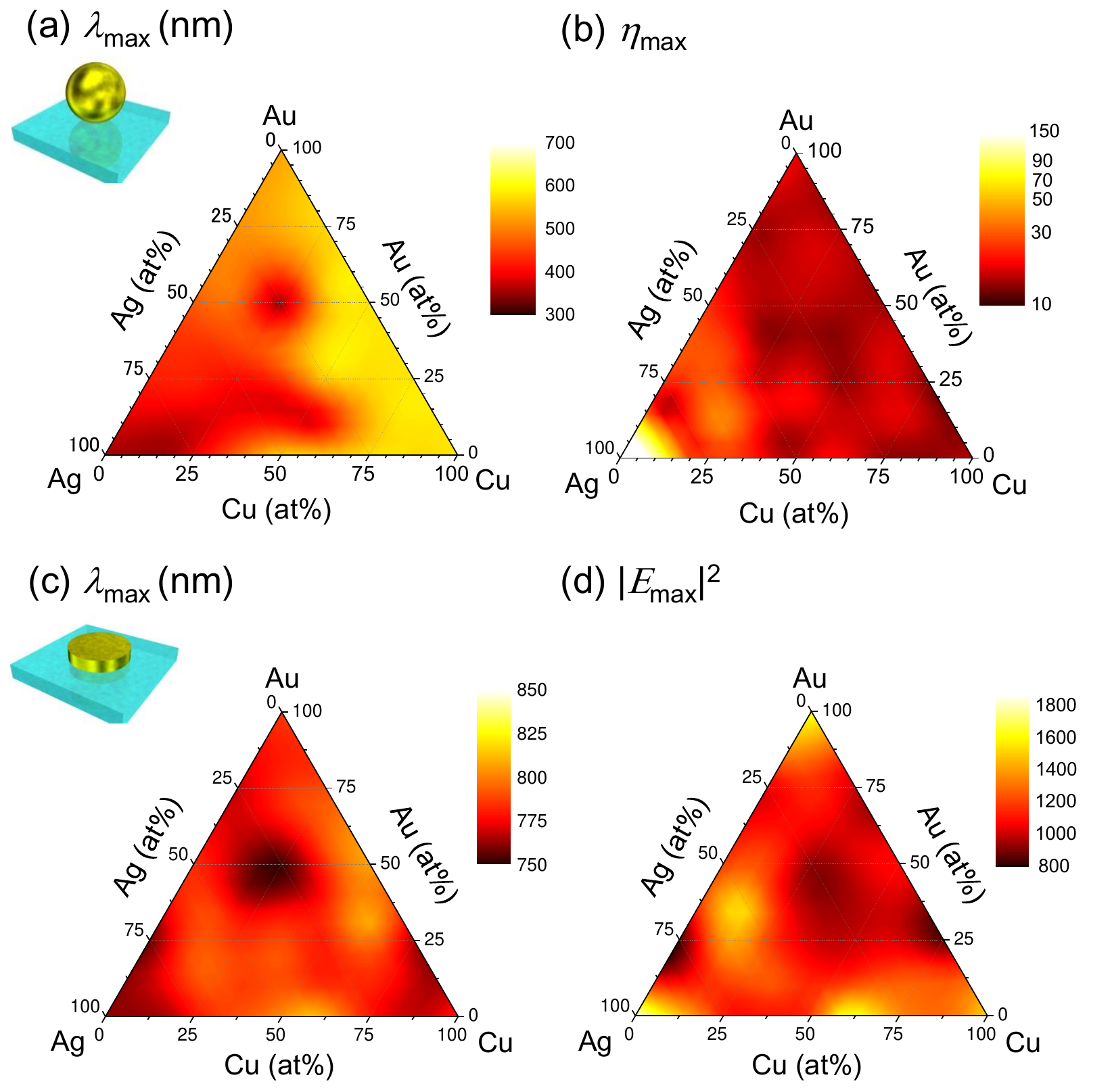} \caption{Summary of
FDTD modeling and numerical calculations. (a) The maximum
wavelength at which the strongest EM intensity was obtained for
nano-spheres. (b) Maximum field intensity for the sphere.  (c)
Wavelength of the plasmon resonance for the nano-disk array. (d)
Maximum $\left|E\right|^2$ for the nano-disk array. The Au:Ag
alloy data were taken from reference~\cite{ynalloy1}.} \label{f3}
\end{figure}

There was a tendency for Au rich alloys to exhibit large $F$
values, most probably due to their higher chemical stability. Ag
and Cu are susceptible to oxidization or/and sulfurization at room
conditions. Permittivity of a given metal alloy is shown to be
strongly affected by the nature of its stoichiometric
constituents, i.e, enrichment in Ag leads to a decreased
$\varepsilon_2$ due to diminished damping of electronic
oscillations typical for silver. Similarly, despite alloying, Au
and Cu rich systems continue to exhibit characteristic interband
absorption related spectral features in the longer wavelength
region. Optical properties of metals are inextricably linked to
the plasmonic response given rise by electrons oscillated due to
the interaction with the EM-field of incident light. In order to
quantitatively evaluate and compare plasmonic properties of
Au:Ag:Cu alloys Drude parameters are extracted to account for the
free electron contribution. Since in the shorter wavelength
UV-visible (VIS) region an optical response is strongly affected
by the interband transitions of bound electrons accounted for by
the Lorenz oscillation terms, Drude parameters were extracted from
experimental data  at $\geq$ 650 nm wavelength. The resulting
values of plasma frequency $\omega_p$ and the damping related
relaxation constant $\tau$ for all 38 different Au:Ag:Cu alloy
compositions are plotted in Figure 3; the values for the Au:Ag
system were determined previously~\cite{ynalloy1}. Plasma
frequency $\omega_p$ of the investigated alloys ranges from 1.2 to
1.6 $\times$ 10$^{16}$ rad/s and directly corresponds to their
free electron densities, which can be expressed as  $\omega_p =
(Ne^{2}/\varepsilon_{0}m)^{1/2}$, where $N$ is the density of
electrons, $e$ is the elementary charge, $m$ is the effective mass
of an electron. Therefore, the observed spectral variations in
different alloys can, to a large extent, be modeled as changes in
$\omega_p$, e.g., an increase in the free carrier density results
in a blue-shift of the plasmon resonance wavelength. For the Au,
Ag, Cu alloy system under consideration, the middle of the area
plot of Figure 3a (delimited by the 75:50:50\% lines for Au:Ag:Cu
respectively) exhibits high plasma frequency values. Conversely,
the other parts, especially ones rich in Ag are shown to have
lower values. Since $\tau$ corresponds to the lifetime of the free
electron oscillation, it directly affects the magnitude of
plasmonic EM field enhancement. While, as expected, pure metals
have large $\tau$ values, it is noteworthy that alloy systems
around Ag (50-60\%) Au (40-30\%) mixed with $\sim$15\% Cu, as well
as at Cu 60\% Ag 40\% seem to also show relatively decreased
damping. Sufficiently narrow confidence intervals of 0.02\% for
the Drude parameters highlight the potential in plasmonic
applications of such alloys. Plasmonic response of alloy
nanoparticles are discussed next.

\subsection*{Plasmonic properties of alloy materials.}

For plasmonic applications of alloys it is of significant
importance how plasmon resonance frequencies and EM field
intensity values relate to their stoichiometric composition. For
this purpose FDTD and numerical simulations of simple plasmonic
structures were performed, using the experimentally determined
complex permittivity values ($\varepsilon_{1} + i\varepsilon_{2}$)
to approximate material properties of Au:Ag:Cu metal alloys. When
conducting numerical analysis of the plasmonic resonance for a
spherical metallic particle in the quasi-static approximation one
can make use of analytical expression for EM field enhancement
$\eta$~\cite{plasmonics}:
\begin {equation}\label{e2}
\eta =  \left| \frac{E}{E_0} \right| ^2 =  \left|
\frac{3\varepsilon}{\varepsilon+2} \right| ^2,
\end{equation}
\noindent where $E_0$ represents the electric field of the
incident light and $E$ is the localized electric field in the
near-field of the plasmonic particle. The ternary plots of
spherical particle resonant wavelengths and their corresponding
maximum values of $\eta$ for various alloy systems are presented
in Figures 4a and 4b, respectively. Equation 1 is only valid for a
sphere, hence, FDTD calculations were carried for nano-discs
(Figure 4c,d). Using alloys, it is possible to tune the plasmonic
resonance over the entire visible spectral range. Such flexibility
is required in applications of color filters and photo-catalysts.
A plasmon wavelength range from 350 to 650 nm was used to monitor
plasmon resonance for the spheres of different elemental
composition. Simulation results match previous experimental and
numerical results~\cite{alloyparticle4, alloyparticle5}. This
wavelength region is strongly affected by the interband
transitions. In the case of nano-disc simulations, a wide range of
plasmon resonances from 750 to 825 nm were obtained. The trend
observed for the composition dependent resonant peak wavelength
shift closely follows results obtained for the spherical
structure, except for the Cu-rich nano-discs. This result
indicates that the plasmon resonance of Cu spheres is red shifted
due to the interband transition of Cu (the Lorenz oscillation).
However, in the case of nano-discs, plasmon resonance is mainly
defined by the Drude contribution and the difference between Au,
Ag, Cu and their alloys become less significant, hence, the
plasmon resonance range becomes narrower.

\begin{figure}[tb]
\centering\includegraphics[width=12cm]{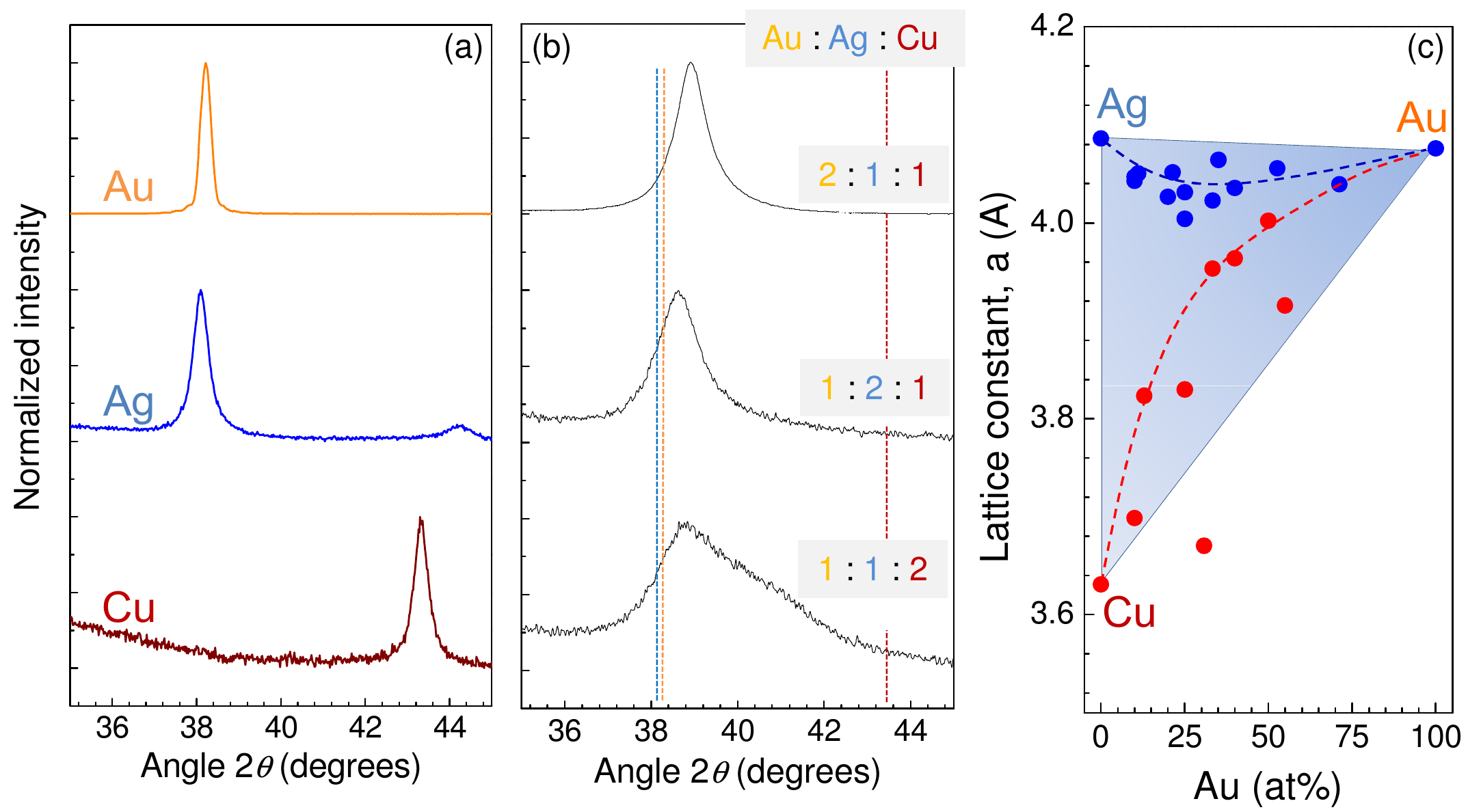} \caption{X-ray
crystal diffractogram of pure Au, Ag and Cu metals (a), and their
alloys at (2:1:1), (1:2:1) and (1:1:2) stoichiometric ratios (b).
(c) Lattice constant, $a$, dependence on the Au concentration in
the alloy system, obtained form the X-ray diffraction angle.
Dashed lines in (c) are guides for the eye.} \label{f4}
\end{figure}

Furthermore, for several alloy systems the EM enhancement was
shown to exceed that of pure Au and Cu. In the case of spheres,
Au:Ag series with 15\% to 5\% of Au, Ag:Cu series with
20$\sim$25\% of Cu, and with 40$\sim$60\% of Ag in Au:Ag:Cu
systems. Similarly high enhancements were observed for nano-discs
in Ag rich compositions (Au 35\%, Ag 52\%, Cu 13\% and Ag 40\% Cu
60\%). The overall EM enhancement trend is consistent with the
relaxation constant $\tau$ of those alloys. For the higher $\tau$
values, electrons experience lower resistance (a high quality
factor $Q$ for the plasmonic resonance) which results in a strong
EM field enhancement.

\subsection*{X-ray crystallography}

To gain further insight into the structural properties of Au-Ag-Cu
alloys and to deduce their influence on the optical response
crystallographic investigation by means of X-ray powder
diffraction measurements has been performed (Figure 5). In the
case of pure metals, diffraction angles 2$\theta$ of around
38$^\circ$ for both Au and Ag, and 43.5$^\circ$ for Cu were
obtained as expected. These values are directly related to their
lattice constants a = 4.078$\AA$ (Au), 4.086$\AA$ (Ag), and
3.615$\AA$ (Cu)~\cite{phasediagram}. Phase diagram for the metal
alloys indicates that at all ratios of composition the Au-Ag
system is arranged in a face centered cubic (fcc) lattice.
Conversely, the Ag-Cu system has an fcc eutectic point at
Ag$_3$Cu$_2$ at the eutectic temperature of 800$^\circ$C, however,
provided splat cooling occurs, a quasi-stable solid solution state
can be obtained in a wide range of compositions. The Au-Cu system
exhibits a super lattice of Cu$_3$Au, CuAu and CuAu$_3$, where
AuCu has a cubic lattice and the two remaining sub-lattices show
the fcc arrangement.

\begin{figure}[tb]
\centering\includegraphics[width=12cm]{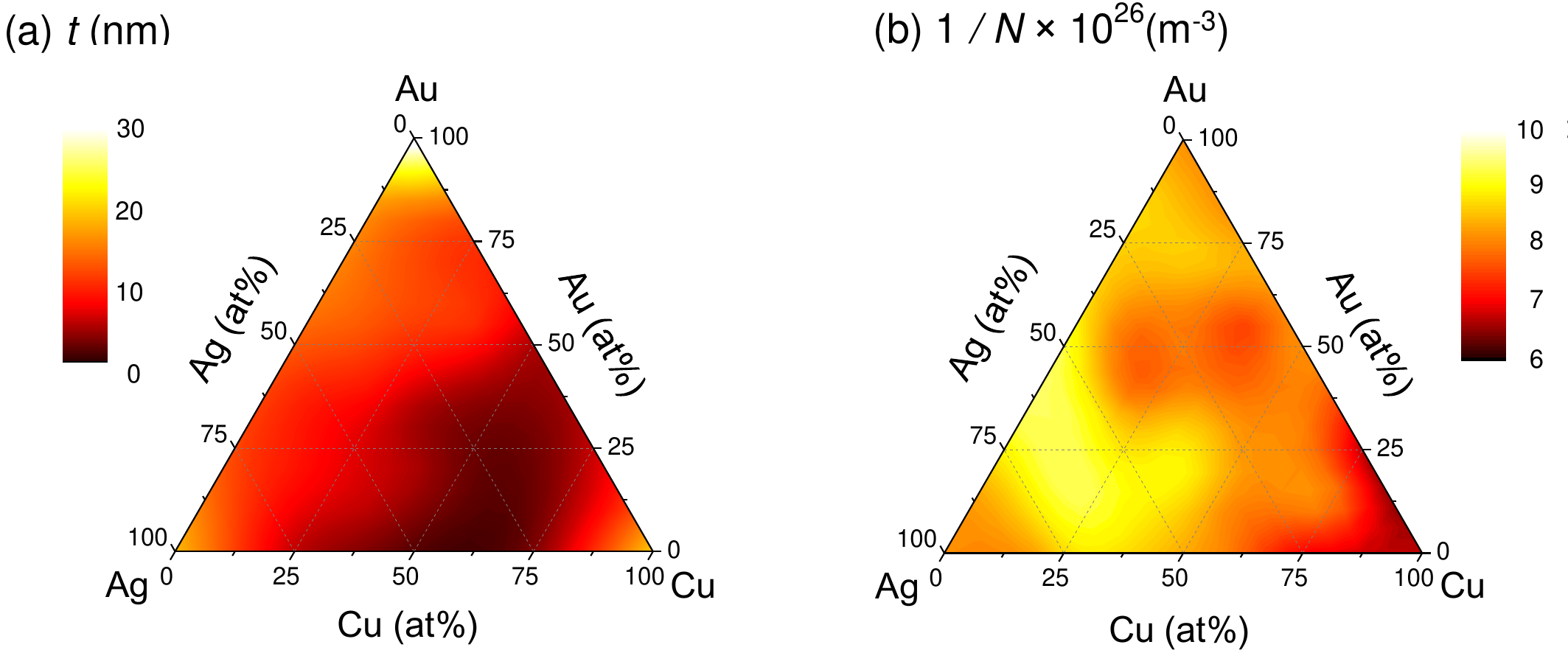} \caption{(a) Ternary
plot of the size, $t$, of crystallites obtained from Scherrer's
analysis of XRD data (Eqn.~\ref{es}). (b) relaxation time which is
reciprocal to the density of scatterers, $1/N$, of the alloys at
varying stoichiometric compositions.} \label{f5}
\end{figure}

In this study, we have focused exclusively on diffraction from the
(111) face, since diffraction peak intensities from other planes
are much weaker, therefore it would be exceedingly difficult to
distinguish between the aforementioned cubic and fcc sub-lattices.
A strong (111) signature indicates that the alloy has a fcc
lattice. With the addition of Cu into the Au-Ag system, the
diffraction peak associated with Au and Ag moved towards larger
2$\theta$ angles. However, at high concentrations of Cu,
diffraction peaks become broader and apparently composed of two
distinct peaks. This can be attributed to the solid solution
formation during the physical vapor deposition step giving rise to
heterogeneity in the metal alloy, which in turn is observed as two
peaks corresponding to different lattice constants. Figure 5 (c)
shows how the lattice constant associated with observed
diffraction peaks depends on the atomic ratio of Au in the alloy.
As the system is enriched in Au the two peaks begin to shift
closer to each other, converging towards the lattice constant of
Au. From this result it can be deduced that Au containing alloys
perfectly intermix, avoiding segregation into separate phases.
Homogeneous crystal phase has lower electric dumping due to
reduced grain boundaries and scattering centers. which favors
larger EM field enhancements as discussed below. Surface enhanced
Raman scattering (SERS) from such alloys with expected high
EM-enhancement is an interesting topic for further investigations.
Chemical attachment of different analytes on the surface of alloy
with different neighboring atoms Ag, Au, or Cu is expected to be
reflected in SERS chemical and EM-enhancement factors.

From the XRD measurement results it is also possible to retrieve
the mean size of crystallites. According to the Scherrer’s
equation, size of crystallites $t$ can be calculated using the
standard expression~\cite{sherrer} :
\begin{equation}\label{es}
t = \frac{K \lambda}{\beta \cos \theta},
\end{equation}
\noindent where $K$ is the dimensionless shape factor in our case
assumed to be 0.9, $\lambda$ is the X-ray wavelength of 0.154 nm
for the Cu-K$\alpha$ line used, $\beta$ is the full width half
maximum (FWHM) of the XRD peak in radians, and $\theta$ is the
Bragg diffraction angle. The crystallite size values extracted
from experimental data are plotted in Figure 6a. The lattice
constant of the fcc structure, $a$, and atom density, $n_{a}$,
were obtained based on the values of the diffraction angle
$\theta$.

Plasmon relaxation constant $\tau$ is the average duration between
two scattering events which define the electron mean free path
(mfp). Therefore, from the density of scatterers we can deduce
$\tau$ according to the following equation:
\begin{equation}
\tau = \frac{1}{(n_{a} + n_{e})\sigma v_F },
\end{equation}
\noindent where $n_{a}$ and $n_{e}$ respectively are the densities
of atoms and electrons, $\sigma$ is the scattering cross-section,
and $v_F$ is the Fermi velocity of electrons. The $n_{e}$ value,
which defines the plasma frequency, can be obtained assuming that,
as is typical for metals, one electron per atom is donated for the
metallic bonding in the alloy. It is noteworthy, that other
scattering mechanisms, e.g., by grain boundaries of
nano-crystallites also contribute to the decrease of the mfp,
i.e., cause a shorter relaxation time $\tau$. The dependence of
scatterer density $1/N = 1/( n_{a} + n_{e})$ on the alloy
composition is plotted in Figure 6b. It shows strong correlation
with the ternary plots of $\tau$ (Figure 3b) obtained from purely
optical measurements. Compositional map of crystallite size $t$
(Figure 6a) can also be seen as representing the relaxation time
component due to the grain boundary derived scattering $\tau
\propto t$.

Mixing of metals in nanoparticles with different surface to volume
ratios can bring new ordering of eutectic metal mixtures as
predicted for the Ag$_6$-Cu$_4$ system, with prevalent formation
of glassy mixtures with an icosahedral ordering around Cu
atoms~\cite{06jns2390}. Alloy films prepared by evaporation as
studied here could serve as a source material to make alloy
nanoparticles using laser ablation via specific formation
mechanisms exploring surface super-cooling/heating~\cite{Farson}
and fragmentation mechanisms\cite{13ijnm2601}. Surface composition
and ordering of alloy atoms is expected to bring new (photo)
catalytic properties and surface functionalization scenarios for
SERS sensing. Laser ablation of alloys using ultra-short laser
pulses which also facilitates an ultra-fast thermal quenching
could open a new availability of amorphous metals for SERS and
photo-catalysis. This is an unexplored area of research since pure
metals are forming crystals but not glasses.

\section*{Conclusions}

The optical permittivity $(\varepsilon_1+i\varepsilon_2)$ of the
plasmonic $d^{10}$-metal Au, Ag and Cu alloy system was determined
for the first time by means of a systematic experimental
investigation over a large number (38 alloys) of intermixing
ratios. The Drude parameters were extracted by numerical analysis
of the optical transmission and reflection spectral measurements
using a high fidelity fit. It is shown that the plasmon resonance
wavelength can be tuned in a wide spectral range by simply
changing the composition of the Au-Ag-Cu metal alloy, providing a
way to engineer the plasma frequency $\omega_{p}$ of the system.
It was confirmed that the plasmon resonance relaxation time,
$\tau$, which governs the EM field enhancement, is dependent on
the crystallinity of the alloy. Strong correlation between $\tau$
values determined by means of XRD and optical measurements is
shown. At certain alloy stoichiometries the EM field enhancement
factor exceeds values observed for pure Au structures. Due to its
resistance to oxidation and sulphurization at atmospheric
conditions, the alloy system has strong potential for a variety of
plasmonic applications including sensing, photo-catalysis, and
solar energy harvesting. New chemical and EM-enhancements in SERS
are expected on surfaces of these plasmonic metal alloys.

\small{\section*{Acknowledgements} YN gratefully thanks Prof.
Toshihiko Baba from Yokohama National University for fruitful
discussions and for granting access to fabrication and
characterization facilities. This work was financially JSPS
Grants-in-Aid for Scientific Research. SJ is grateful for support
via Australian Research Council DP130101205 project and startup
funding for Nanolab by Swinburne University.}

\small{\section*{Author contributions statement} Y.N. conceived
the experiment,  Y.H. and Y.N. conducted the experiments, Y.N.,
A.B., G.S. and S.J. analysed data and edited the manuscript. All
authors reviewed the manuscript.}



\textbf{Competing financial interests}: The authors declare no
competing financial interests.


\end{document}